\documentclass[aps,preprint,superscriptaddress]{revtex4}
\usepackage{amssymb}
\usepackage{amsmath}
\usepackage{graphicx}
\usepackage{fancyhdr}
\newcommand{\bea}{\begin{eqnarray}}
\newcommand{\eea}{\end{eqnarray}}

\begin{document}

\title{On duality of the noncommutative supersymmetric
  Maxwell-Chern-Simons theory}

\author{M. Gomes}
\email{mgomes@fma.if.usp.br}
\affiliation{Instituto de F\'\i sica, Universidade de S\~ao Paulo\\
Caixa Postal 66318, 05315-970, S\~ao Paulo, SP, Brazil}

\author{J. R. Nascimento}
\email{jroberto@fisica.ufpb.br}
\affiliation{{ Departamento de
F\'{\i}sica, Universidade Federal da Para\'\i ba, Caixa Postal 5008, 58051-970,
Jo\~ao Pessoa, PB, Brasil}}
\affiliation{Instituto de F\'\i sica, Universidade de S\~ao Paulo\\
Caixa Postal 66318, 05315-970, S\~ao Paulo, SP, Brazil}

\author{A. Yu. Petrov}
\email{petrov@fisica.ufpb.br}
\affiliation{{ Departamento de
F\'{\i}sica, Universidade Federal da Para\'\i ba, Caixa Postal 5008, 58051-970,
Jo\~ao Pessoa, PB, Brasil}}

\author{A. J. da Silva}
\email{ajsilva@fma.if.usp.br}
\affiliation{Instituto de F\'\i sica, Universidade de S\~ao Paulo\\
Caixa Postal 66318, 05315-970, S\~ao Paulo, SP, Brazil}

\author{E. O. Silva}
\email{edilberto@fisica.ufpb.br}
\affiliation{{ Departamento de
F\'{\i}sica, Universidade Federal da Para\'\i ba, Caixa Postal 5008, 58051-970,
Jo\~ao Pessoa, PB, Brasil}}

\date{\today}

\begin{abstract}

We study the possibility of establishing the dual equivalence between the noncommutative supersymmetric Maxwell-Chern-Simons theory and the noncommutative supersymmetric self-dual theory. It turns to be that whereas in the commutative case the Maxwell-Chern-Simons theory can be mapped into the sum of the self-dual theory and the Chern-Simons theory, in the noncommutative case such a mapping is possible only for the theory with modified Maxwell term.
 
\end{abstract}

%\keywords{equivalence, susy self-dual, susy maxwell-chern-simons}

\maketitle


\section{Introduction}
The duality, allowing to construct mappings between different field
theories is a very important aspect of three-dimensional
field models. Initially the duality was observed for the
example of the free Maxwell-Chern-Simons and self-dual theories
\cite{nieu}. Further, in a number of papers \cite{many} different
methods of implementing the duality were studied. The development of
 noncommutative field theories brought a question about possible
generalization of duality in this situation. There, the ordering
problem of product of fields turns out to be fundamental, at least in
the  application of
the gauge embedding method \cite{our}. 

One approach of implementing the duality for the noncommutative field
theories is based on the use of the Seiberg-Witten map, as it was
developed in \cite{SW}. We would like to point out that there is an
alternative method of construction of dual models for the
noncommutative theories which has been previously developed in
\cite{embed} and successfuly applied to the study of the duality. This
method is based on an appropriate change of variables allowing to
rewrite the action in a simpler form, with a decreased  number
of derivatives. As a result, the modified
Maxwell-Chern-Simons action turns out to be mapped into two theories,
one of them being the Chern-Simons theory whereas the other one is the
self-dual model.
Here, our aim is to study this method in the noncommutative case within
the framework of the superfield formulation of supersymmetric field theories.
By conveniently deforming the original Lagrangian in the Wess-Zumino gauge, we
demonstrate that  the mentioned duality holds in the physical sector.
\section{The Maxwell-Chern-Simons theory}
The starting point of our study is the Maxwell-Chern-Simons theory
whose action looks like (we follow the notations of~\cite{SGRS,MCS}):
\begin{eqnarray}
\label{action}
S&=&\frac{1}{2g^2}\int d^5 z \Big[-\frac{1}{2}W^\alpha
*W_\alpha+m\Big( A^{\alpha}
*W_{\alpha}+\nonumber\\&+&\frac{i}{6}\{A^{\alpha},
A^{\beta}\}_**D_{\beta}A_{\alpha}
+\frac{1}{12}\{A^{\alpha},A^{\beta}\}_**
\{A_{\alpha},A_{\beta}\}_*\Big)\Big]\,.  \label{2n},
\end{eqnarray}
where
\begin{eqnarray}
\label{sstr}
W_\beta =\frac{1}{2}D^\alpha D_\beta A_\alpha -
\frac{i}{2}[A^\alpha ,D_\alpha A_\beta ]_*-
\frac{1}{6}
[A^\alpha ,\{A_\alpha ,A_\beta \}_*]_*
\end{eqnarray}
is the superfield strength constructed on the base of the the spinor
superpotential $A_\alpha$. In Eq.~(\ref{action}), the first and the
second terms are the noncommutative Maxwell and Chern-Simons terms,
respectively. Due to the noncommutativity, this action, although
Abelian,  includes the  self-interactions for the gauge
superfield. The  parameter $m$ is the topologica; mass of the
superfield.  Hereafter it is implicitly assumed that all commutators
and anticommutators are Moyal ones, that is,
$[A,B\}~\equiv~A~*~B~\mp~B~*~A$,  with
\begin{equation}
A(x)*B(x)\,\equiv\, A(x)\,
exp\left(\frac{i}{2}\,\overleftarrow{\frac{\partial}{\partial
x^{\mu}}}\,\Theta^{\mu\nu}\,\overrightarrow{\frac{\partial}{\partial
x^{\nu}}}\right)\,B(x),
\end{equation}
being the Moyal-Groenewald $*$-product.

\section{Dual projecting of the theory}
Let us carry the dual projection of the above theory. To do it, we
introduce  the auxiliary spinor superfield $\pi^{\alpha}$ to
lower the order of the Lagrangian, that is, the number of interacting
fields in vertices.
With this objective, it is natural to suggest that
the  equivalent form of this action is
\bea
\label{equiv}
S&=&\frac{1}{2g^2}\int
d^5z\Big(\frac{k_1}{2}(\pi^{\alpha}-A^{\alpha})*(\pi_{\alpha}-A_{\alpha})+k_2
\pi^{\alpha}D^{\beta}D_{\alpha}A_{\beta}+
k_3\pi^{\alpha}*[A^{\beta},D_{\beta}A_{\alpha}]+\nonumber\\&+&
%k_4\pi^{\alpha}[A^{\beta},\{A_{\alpha},A_{\beta}\}]+\nonumber\\&+&
l_1\{A^{\alpha},A^{\beta}\}*D_{\beta}A_{\alpha}
%+l_2\{A^{\alpha},A^{\beta}\}*\{A_{\alpha},A_{\beta}\}
\Big).
\eea
 To simplify the situation, we employ the
  Wess-Zumino gauge \cite{SGRS} in which the (Moyal) products  of
  three and more spinor superfields which are not affected by the action of
  derivatives, as f.e. $A_{\alpha}*A_{\beta}*A_{\gamma}$,
   are equal to zero.
Here $k_1,k_2,k_3, l_1,$ are constants to be fixed. The equation
of  motion for the $\pi_{\alpha}$ is
\bea
\pi_{\alpha}=A_{\alpha}-\frac{k_2}{k_1}D^{\beta}D_{\alpha}A_{\beta}-\frac{k_3}{
  k_1}[A^{\beta},D_{\beta}A_{\alpha}].
\eea
By substituting this $\pi_{\alpha}$ into the action (\ref{equiv}), we arrive at
\bea
S&=&\frac{1}{2g^2}\int d^5 z\Big[-\frac{k^2_2}{2k_1}
D^{\gamma}D^\alpha A_{\gamma}D^{\beta}D_{\alpha}A_{\beta}
-\frac{k_2k_3}{k_1}D^{\gamma}D^\alpha
A_{\gamma}*
[A^\beta ,D_\beta A_\alpha ]-\nonumber\\&-&
\frac{k^2_3}{2k_1}[A^{\gamma},D_{\gamma}A^\alpha ]*[A^\beta ,D_\beta A_\alpha
]+
%\nonumber\\&+&
(k_2A^{\alpha}D^{\gamma}D_{\alpha}A_{\gamma}+(k_3+l_1)\{A^{\alpha},A^{\beta}\}*
D_{\beta}A_{\alpha}
%\nonumber\\&+&(k_4+l_2)\{A^{\alpha},A^{\beta}\}*\{A_{\alpha},A_{\beta}\}
\Big)\Big]\,,\nonumber
\eea
whereas the expanded form of the Maxwell-Chern-Simons action in the
Wess-Zumino gauge is (cf. \cite{MCS})
\begin{eqnarray}
\label{sint}
S&=&\frac{1}{2g^2}\int d^5 z\Big[-\frac{1}{8}
D^{\gamma}D^{\alpha} A_{\gamma}D^{\beta}D_{\alpha}A_{\beta}
+\frac{i}{4}D^{\gamma}D^{\alpha}
A_{\gamma}*
[A^\beta ,D_\beta A_\alpha ]+\nonumber\\&+&
%\frac{1}{12}D^{\gamma}D^\alpha A_{\gamma}*[A^\beta ,\{A_\beta ,A_\alpha \}]+
%\nonumber\\&+&
\frac{1}{8}[A^{\gamma},D_{\gamma}A^\alpha ]*[A^\beta ,D_\beta A_\alpha
]%-\frac{i}{12}
%[A^{\gamma},D_{\gamma}A^\alpha ]*[A^\beta ,\{A_\beta ,A_\alpha \}]-
%\nonumber\\&-&\frac{1}{72}
%[A^{\gamma},\{A_{\gamma},A^\alpha \}]*[A^\beta ,\{A_\beta ,A_\alpha\}]
+%\nonumber\\ &+&
m\Big(\frac{1}{2}A^{\alpha}D^{\beta}D_{\alpha}A_{\beta}-\frac{i}{3}\{A^{\alpha},
A^{\beta}\}*D_{\beta}A_{\alpha}
%-\frac{1}{12}\{A^{\alpha},A^{\beta}\}*\{A_{\alpha},A_{\beta}\}
\Big)\Big]\,.
\end{eqnarray}
Comparing the above expressions we obtain
$k_1=m^2$, $k_2=\frac{m}{2}$ (which is easily found in the commutative
case). Further, $k_3=-\frac{im}{2}$, 
$l_1=\frac{im}{6}$. Thus, the action
(\ref{equiv}) is found to be:
\bea
\label{equiv1}
S&=&\frac{1}{2g^2}\int
d^5z\Big(\frac{m^2}{2}(\pi^{\alpha}-A^{\alpha})*
(\pi_{\alpha}-A_{\alpha})+\frac{m}{2}\pi^{\alpha}D^{\beta}D_{\alpha}A_{\beta}-
\frac{im}{2}\pi^{\alpha}*[A^{\beta},D_{\beta}A_{\alpha}]+
%\nonumber\\&-&\frac{m}{6}\pi^{\alpha}*[A^{\beta},\{A_{\alpha},A_{\beta}\}]+
\nonumber\\&+&
\frac{im}{6}A^{\alpha}*[A^{\beta},D_{\beta}A_{\alpha}]
%+\frac{m}{12}\{A^{\alpha},A^{\beta}\}*\{A_{\alpha},A_{\beta}\}
\Big).
\eea
Now, we introduce $\pi_{\alpha}=f^+_{\alpha}+f^-_{\alpha}$,
$A_{\alpha}=f^+_{\alpha}-f^-_{\alpha}$, so that the quadratic part of this action takes the form
\bea
\label{quadr}
S_2=\frac{1}{2g^2}\int
d^5z(2m^2f^{-\alpha}f^-_{\alpha}-\frac{m}{2}f^{-\alpha}D^{\beta}D_{\alpha}
f^-_{\beta}+
\frac{m}{2}f^{+\alpha}D^{\beta}D_{\alpha}f^+_{\beta}),
\eea
which is a sum of the quadratic part of the self-dual action for the
$f^-_{\alpha}$ field and the quadratic part of the Chern-Simons action
for the $f^+_{\alpha}$ field. The interaction part, however, is much
more complicated than in \cite{embed}. Note, however, that it involves
terms only up to fourth order in the fields whereas the original
Maxwell-Chern-Simons action involves terms up to the sixth order.

The vertex of third order in the fields in the
action  (\ref{equiv1}) looks like
\bea
V_3=-i\frac{m}{3}\int d^5
z(f^{+\alpha}+2f^{-\alpha})*[f^{+\beta}-f^{-\beta},
D_{\beta}f^+_{\alpha}- D_{\beta}f^-_{\alpha}],
\eea
from which we see that the $f^{+\alpha}$ gives the Chern-Simons triple
term with the correct coefficient (that is, $-\frac{i}{3}$), but
$f^{-\alpha}$ with  a wrong one (that is, $-\frac{2i}{3}$). A similar
situation  was observed in \cite{embed}. We note also the presence of
"mixed" terms.

From this result, we conclude that the noncommutativity destroys
duality  in the "pure"
sense.  To evade this situation, we introduce a deformed action in a
way 
similar to \cite{embed}, 
\bea
\label{equiv2}
S_1&=&\frac{1}{2g^2}\int
d^5z\Big(\frac{m^2}{2}(\pi^{\alpha}-A^{\alpha})*
(M^{-1})_{\alpha\beta}*
(\pi^{\beta}-A^{\beta})+\frac{m}{2}\pi^{\alpha}
D^{\beta}D_{\alpha}A_{\beta}- \nonumber\\&-& \frac{im}{2}\pi^{\alpha}*
[A^{\beta},D_{\beta}A_{\alpha}]+
\frac{im}{6}A^{\alpha}*[A^{\beta},D_{\beta}A_{\alpha}]
\Big),
\eea
where $(M^{-1})_{\alpha\beta}$ is a matrix to be  determined.  The
corresponding  deformed Maxwell-Chern-Simons action is
\bea
S&=&\frac{1}{2g^2}\int d^5 z \Big[-\frac{1}{2}W^\alpha
*M_{\alpha\beta}* W^{\beta}+m\Big( A^{\alpha}*W_{\alpha}+
%\nonumber\\&+&
\frac{i}{6}\{A^{\alpha},A^{\beta}\}_**D_{\beta}A_{\alpha}\Big)\Big]\,,
\eea
where the $W_{\alpha}$ is restricted in the Wess-Zumino gauge by first
two terms of (2).
It is clear that the matrix $M^{-1}$ should be of the form
\bea
(M^{-1})_{\alpha\beta}=-C_{\alpha\beta}+\Lambda_{\alpha\beta}[f],
\eea
with $\Lambda_{\alpha\beta}[f]|_{f_{\alpha}=0}=0$.

Our aim is to fix the $\Lambda_{\alpha\beta}[f]$ in a way 
allowing for the arisal of the Chern-Simons and self-dual actions, that
is, we want to obtain $S_1$ in the form
\bea
\label{aim}
S_1&=&\frac{1}{2g^2}\int d^5 z \Big[2m^2f^{-\alpha}f^-_{\alpha}-
%\nonumber\\&-&
\frac{m}{2}f^{-\alpha}D^{\beta}D_{\alpha}f^-_{\beta}+\frac{im}{3}\{f^{-\alpha},f^{-\beta}\}_**D_{\beta}f^-_{\alpha}+\\&+&
\frac{1}{2g^2}\int d^5 z \Big[\frac{m}{2}f^{+\alpha}D^{\beta}D_{\alpha}f^+_{\beta}-\frac{im}{3}\{f^{+\alpha},f^{+\beta}\}_**D_{\beta}f^+_{\alpha}
\Big].\nonumber
\eea
We note that the quadratic part of this action already was
obtained in Eq. (\ref{quadr}), so, it remains to fix the triple and
quartic terms.  This can be done via the undetermined coefficients
method. So, the
problem is  to choose $\Lambda_{\alpha\beta}$ to satisfy the relation
\bea
\label{therel}
&&2m^2f^{-\alpha}*\Lambda_{\alpha\beta}*f^{-\beta}-\frac{im}{3}\{f^{+\alpha}+
2f^{-\alpha},f^{+\beta}-f^{-\beta}\}*(D_{\beta}f^+_{\alpha}-D_{\beta}
f^-_{\alpha})%-\nonumber\\&-&
%\frac{m}{12}\{f^{+\alpha}+3f^{-\alpha},f^{+\beta}-f^{-\beta}\}*\{f^+_{\alpha}-
%f^-_{\alpha},f^+_{\beta}-f^-_{\beta}\}
=\nonumber\\&=&
\frac{im}{3}\{f^{-\alpha},f^{-\beta}\}_**D_{\alpha}f^-_{\beta}
%+\frac{m}{12}\{f^{-\alpha},f^{-\beta}\}_**\{f^-_{\alpha},f^-_{\beta}\}_*-
%\nonumber\\&-&
-\frac{im}{3}\{f^{+\alpha},f^{+\beta}\}_**D_{\alpha}f^+_{\beta}
%\frac{m}{12}\{f^{+\alpha},f^{+\beta}\}_**\{f^+_{\alpha},f^+_{\beta}\}_*
.
\eea
It is easy to check that the terms involving only $f^+$ fields in the
left-  and the right-hand sides of this equation exactly coincide. The
terms with two or more $f^-$ fields must be cancelled by the
term  $f^{-\alpha}*\Lambda_{\alpha\beta}*f^{-\beta}$. 
The only remaining difficulty is related to the
terms with  only one $f^-$ field (all other fields carry $+$
signs). However, the sum of these
"dangerous" terms vanishes. In fact, for triple terms we have
\bea
\int d^5 z\left[-\frac{2im}{3}\{f^{-\alpha},f^{+\beta}\}*D_{\beta}f^+_{\alpha}+
\frac{im}{3}\{f^{+\alpha},f^{-\beta}\}*D_{\beta}f^+_{\alpha}+
\frac{im}{3}\{f^{+\alpha},f^{+\beta}\}*D_{\beta}f^-_{\alpha}\right],
\eea
or, in a more explicit form
\bea
&&\frac{2m}{3}\int
d^2\theta\int\frac{d^3k_1d^3k_2d^3k_3}{(2\pi)^9}(2\pi)^3
\delta(k_1+k_2+k_3) \sin(k_2\wedge k_3)\times\\&\times&
\Big(2f^{-\alpha}(k_1)f^{+\beta}(k_2)D_{\beta}f^+_{\alpha}(k_3)
-f^{+\alpha}(k_1)f^{-\beta}(k_2)D_{\beta}f^+_{\alpha}(k_3)-
f^{+\alpha}(k_1)f^{+\beta}(k_2)D_{\beta}f^-_{\alpha}(k_3)\Big).\nonumber
\eea
After integration by
parts  this expression can be rewritten as
\bea
\label{this}
&&\frac{2m}{3}\int d^2\theta 
\int\frac{d^3k_1d^3k_2d^3k_3}{(2\pi)^9}(2\pi)^3\delta(k_1+k_2+k_3)
\sin(k_2\wedge k_3)\times\\&\times&
\Big(2f^{-\alpha}(k_1)f^{+\beta}(k_2)D_{\beta}f^+_{\alpha}(k_3)
-f^{+\alpha}(k_1)f^{-\beta}(k_2)D_{\beta}f^+_{\alpha}(k_3)-\nonumber\\
&-&
f^{+\alpha}(k_1)D_{\beta}f^{+\beta}(k_2)f^-_{\alpha}(k_3)
+D_{\beta}f^{+\alpha}(k_1)f^{+\beta}(k_2)f^-_{\alpha}(k_3)\Big).\nonumber
\eea
After relabelling indices, the last
term in the parenteses of this expression takes the form:
$-f^{-\alpha}(k_1)f^{+\beta}(k_2)D_{\beta}f^+_{\alpha}(k_3)$,
and the whole Eq. (\ref{this}) is rewritten as
\bea
&&\frac{2m}{3}\int d^2\theta\int\frac{d^3k_1d^3k_2d^3k_3}{(2\pi)^9}
(2\pi)^3\delta(k_1+k_2+k_3)\sin(k_2\wedge k_3)\times\\&\times&
\Big[f^{-\alpha}(k_1)f^{+\beta}(k_2)
\Big(D_{\beta}f^+_{\alpha}(k_3)-D_{\alpha}f^+_{\beta}(k_3)\Big)-
f^{+\alpha}(k_1)f^-_{\alpha}(k_2)D^{\gamma}f^+_{\gamma}(k_3)\Big].\nonumber
\eea
Taking into account that
$D_{\beta}f^+_{\alpha}-D_{\alpha}f^+_{\beta}=C_{\alpha\beta}
D^{\gamma}f^+_{\gamma}$, 
we find that this term identically vanishes. 

After carrying out simplifications in the remaining terms, suggesting
that
$\Lambda_{\alpha\beta}$ of
first order in the $f^{\pm\alpha}$ fields, we find
\bea
\label{there21}
&&\int
d^5z\Big[2mf^{-\alpha}*\Lambda_{\alpha\beta}*f^{-\beta}+\frac{i}{3}(2\{f^{-\alpha},f^{+\beta}\}-\{f^{+\alpha},f^{-\beta}\})*D_{\beta}f^-_{\alpha}
+\nonumber\\ &+&
\frac{2i}{3}\{f^{-\alpha},f^{-\beta}\}*D_{\beta}f^+_{\alpha}\Big]
=%\nonumber\\&=&
i\int d^5 z\{f^{-\alpha},f^{-\beta}\}_**D_{\beta}f^-_{\alpha}.
\eea
From this equation one can find
$\Lambda_{\alpha\beta}$ (which
depends on  phase factors). 

 First, one can write down a more explicit form of (\ref{there21}):
\bea
\label{triple}
&&\int d^2\theta
\int\frac{d^3k_1d^3k_2d^3k_3}{(2\pi)^9}(2\pi)^3\delta(k_1+k_2+
k_3)\Big[2mf^{-\alpha}(k_1)\Lambda_{\alpha\beta}(k_2)
f^{-\beta}(k_3)- \nonumber\\ &-&\frac{1}{3}\left[4\sin(k_1\wedge
k_2)f^{-\alpha}(k_1)f^{+\beta} (k_2)-
2\sin(k_1\wedge
k_2)f^{+\alpha}(k_1)f^{-\beta}(k_2))\right]D_{\beta}f^-_{\alpha}
(k_3)-\nonumber\\ &-&\frac{4}{3}
\sin(k_1\wedge k_2)f^{-\alpha}(k_1)f^{-\beta}(k_2)D_{\beta}f^+_{\alpha}(k_3)\Big]
=\\ &=&-2\int d^2\theta
\int\frac{d^3k_1d^3k_2d^3k_3}{(2\pi)^9}(2\pi)^3\delta(k_1+k_2+
k_3)\sin(k_1\wedge k_2)f^{-\alpha}(k_1)f^{-\beta}(k_2)
D_{\beta}f^-_{\alpha}(k_3).\nonumber
\eea
By comparing of the left- and right-hand sides of Eq. (\ref{triple})
we find:
\bea
&&\int d^5 z
f^{-\alpha}*\Lambda_{\alpha\beta}[f]*f^{-\beta}=-(2\pi)^3 
\sin(k_1\wedge k_2)\delta(k_1+k_2+k_3)f^{-\alpha}(k_1)\times\\ &\times&
\Big[-\frac{1}{2m}[D_{\alpha}f^-_{\beta}(k_2)+D_{\beta}f^-_{\alpha}(k_2)]+
\nonumber\\&+&
\frac{1}{3m}\big(-f^+_{\beta}(k_2)D_{\alpha}-
2D_{\beta}f^+_{\alpha}(k_2)+2C_{\alpha\beta}f^{+\gamma}(k_2)D_{\gamma}\big)\Big]
f^{-\beta}(k_3),\nonumber
\eea
which within the expression (\ref{there21}) looks like:
\bea
\Lambda_{\alpha\beta}[f]&=&\frac{i}{2m}(D_{\alpha}f^-_{\beta}+
D_{\beta}f^-_{\alpha})-\frac{i}{3m}(-f^+_{\beta}D_{\alpha}-
2D_{\beta}f^+_{\alpha}+2C_{\alpha\beta}f^{+\gamma}D_{\gamma}).
\eea

So, the manifest form of $\Lambda_{\alpha\beta}$ was found. Thus the dual projection of
the  modified noncommutative Maxwell-Chern-Simons theory was constructed.

\section{Summary}

We have  succeeded in  mapping the noncommutative supersymmetric
Maxwell-Chern-Simons  theory with the modified Maxwell term into the
sum of  the noncommutative Chern-Simons theory and noncommutative
self-dual  theory in the Wess-Zumino gauge. The essential result is that to achieve this
mapping we  must modify the Maxwell term introducing the nontrivial
matrix  $M_{\alpha\beta}$. The appearance of this matrix is a natural
implication of  noncommutativity (and, thus, of a nontrivial
self-interaction).  In principle, this modification can be treated as
some  nonlinear extension of the initial Maxwell-Chern-Simons theory.

{\bf Acknowledgments.}
This work was partially supported by Funda\c c\~ao de Amparo \`a
Pesquisa do  Estado de S\~ao Paulo (FAPESP) and Conselho Nacional de
Desenvolvimento Cient\'\i fico e Tecnol\'ogico (CNPq). The work by
A. Yu. P. has been supported by CNPq-FAPESQ DCR program, CNPq project
No. 350400/2005-9.

\end{document}